\documentclass[aps,prl,showpacs,reprint]{revtex4-1}
\usepackage{bm,epsfig,graphics,amssymb,amsmath,subeqnarray,setspace,graphicx,amsthm,epstopdf,subfigure,color,txfonts}
\usepackage{setspace,natbib}

\usepackage{epsfig,graphics,amssymb,amsmath,color,setspace}
\usepackage{graphicx}
\usepackage{natbib}
\usepackage[sfdefault=cmbr]{isomath}

\def\t{\tensorsym}

\newcommand{\ud}[1]{\overset{\triangledown}{#1}}

\def\b{\bm}

\definecolor{DarkBlue}{rgb}{0,0,0.0}

\newcommand\tcb{\textcolor{DarkBlue}}

\usepackage{graphicx,color,bm}
\graphicspath{{converted_graphics/}}
\usepackage{graphicx}


\begin{document}

\title{Locomotion of helical bodies in viscoelastic fluids: enhanced swimming at large helical amplitudes}
\author{Saverio E. Spagnolie}
\email{spagnolie@math.wisc.edu}
\affiliation{Department of Mathematics, University of Wisconsin, 480 Lincoln Drive, Madison, WI 53706, USA}
\author{Bin Liu}
\affiliation{School of Engineering, Brown University, 182 Hope Street, Providence, RI 02912, USA}
\author{Thomas R. Powers}
\email{thomas\_powers@brown.edu}
\affiliation{Department of Physics and School of Engineering, Brown University, 182 Hope Street, Providence, RI 02912, USA}

\date{\today}

\begin{abstract}
The motion of a rotating helical body in a viscoelastic fluid is considered. In the case of force-free swimming, the introduction of viscoelasticity can either enhance or retard the swimming speed and locomotive efficiency, depending on the body geometry, fluid properties, and the body rotation rate. Numerical solutions of the Oldroyd-B equations show how previous theoretical predictions break down with increasing helical radius or with decreasing filament thickness. Helices of large pitch angle show an increase in swimming speed to a local maximum at a Deborah number of order unity. The numerical results show how the small-amplitude theoretical calculations connect smoothly to the large-amplitude experimental measurements.
\end{abstract}

\maketitle

Much has been learned about the swimming of microorganisms in viscous environments over the last decade~\cite{lp09}. The peculiar behavior of complex fluids has also seen a recent burst of renewed interest, particularly as applied to biological systems~\cite{sm10}. Progress in both fields has begun to blur together, since many organisms commonly swim in shear-thinning or viscoelastic fluids. Some of those fluids are complex specifically because of suspensions of microorganisms swimming and diffusing throughout~\cite{pk92,ip07,slp10,Saintillan10,foffano2012bulk}. Examples of microorganisms swimming in complex fluid environments include mammalian spermatozoa through cervical fluid~\cite{sp06}, the Lyme disease spirochete {\it B. burgdorferi} through the extracellular matrix of our skin~\cite{ks90,hdcbbfrw12}, and the nematode {\it C. elegans} in water-saturated soil~\cite{Jung10}. Organisms such as {\it H. pylori} have even been found to reduce fluid elasticity in order to swim through mucus~\cite{celli09}. 

A puzzle has recently emerged in the study of swimming through complex fluids. Theory, experiment, and simulation have indicated the possibility of both enhancement and retardation of swimming speeds in viscoelastic fluids (see Fig.~1a-f). Helically-shaped bacteria such as {\it Leptospira} and {\it B. burgdorferi} swim faster in solutions with methylcellulose than in non-viscoelastic solutions of the same viscosity~\cite{bt79,ks90}. {\it C. elegans}, however, which propels itself by planar body undulations, swims slower in a viscoelastic fluid than in a viscous fluid~\cite{sa11}. Spermatozoan cells swim slower when the fluid has an elastic response, but along straighter paths due to resultant changes in the flagellar shape, with hyperactivated spermatozoan cells swimming faster than normal cells~\cite{sd92}. The consequences of fluid viscoelasticity on swimming is not, then, a question that can be answered broadly; effects appear to depend sensitively upon the geometry of the swimming stroke and the rheology of the complex fluid.
 
There have also been a number of recent analytical, numerical, and scale-model explorations. Analysis is commonly performed on the Oldroyd-B equations, which describe a viscoelastic flow with no shear-thinning or thickening~\cite{bah87,Larson99}. Using the Oldroyd-B model and others, Lauga showed that an infinite sheet passing small amplitude waves always swims slower with the introduction of viscoelasticity~\cite{Lauga07}. An identical factor of swimming speed reduction was recovered by Fu et al. for a nearly cylindrical body of small pitch angle when passing helical waves~\cite{fwp09}, and similarly for the passage of planar waves~\cite{fpw07}. Teran et al. showed that finite undulatory bodies of large wave amplitude can swim faster in a viscoelastic fluid~\cite{tfs10}, while Curtis and Gaffney showed the same for a three-sphere swimmer \cite{cg13}\tcb{, as did Espinosa-Garcia et al. for flexible swimmers~\cite{elz13}}. Finally, Liu et al. studied experimentally the motion of a rotating, force-free helical filament in a (viscoelastic) Boger fluid, finding that the swimming speed increased or decreased with viscoelasticity depending on the body geometry and rotation rate (see Fig.~1f)~\cite{lpb11,dlfbbpk13}.

In this paper we bridge the gap between the analytical predictions and the experimental and numerical observations just described. By studying numerically the swimming of a helical body in an Oldroyd-B model fluid, we show that the theoretical efforts do indeed capture the effects of viscoelasticity when the helical pitch angle is small and the filament radius is large: namely, that the swimming speed of a rotating helical body in this regime is always smaller than the same in a Newtonian fluid. We will then show how these theories break down for helices of large pitch angle and that the swimming speed can increase with the introduction of viscoelasticity. The results may improve our understanding of mammalian fertility and the spread of bacterial infections and diseases~\cite{hdcbbfrw12}.

\begin{figure*}
\begin{center}
\includegraphics[width=7in]{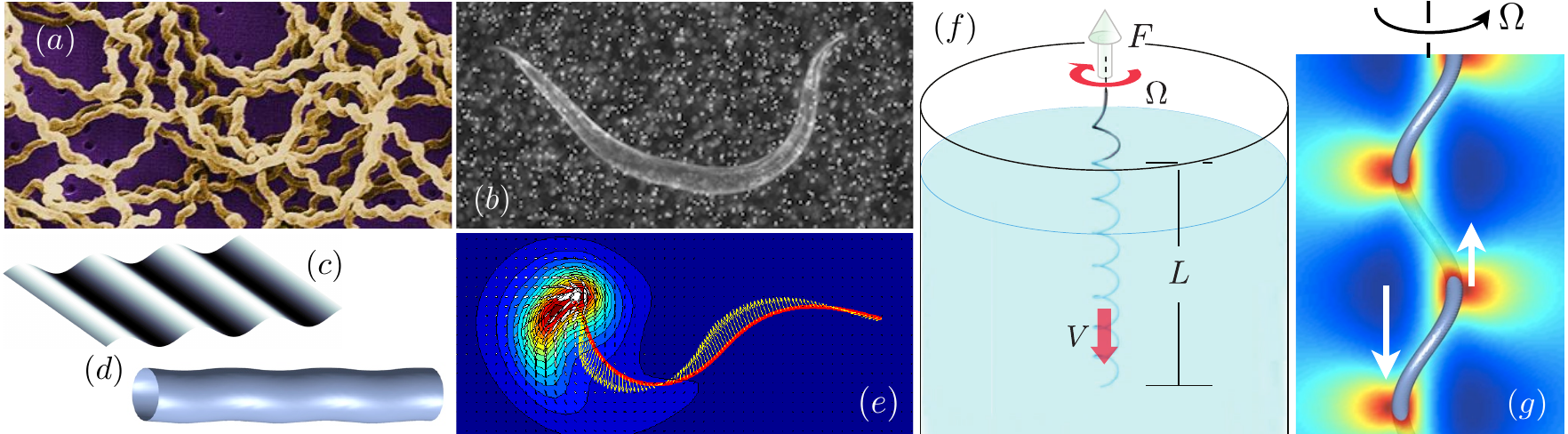}
\caption{(Color online) Experiments, theories, and simulations of swimming in viscoelastic fluids. With increasing fluid elasticity, (a) the helical bacterium {\it Leptospira} swims faster (see also Fig.~7 of ~\cite{bharti03})~\cite{CDC,bt79,ks90}; (b) the nematode {\it C. elegans} swims slower \cite{sa11,saFig}; (c) a two-dimensional swimming sheet of small wave amplitude swims slower~\cite{Lauga07}; (d) a nearly-cylindrical swimmer passing helical waves of small pitch angle swims at the same speed as a two-dimensional sheet~\cite{fwp09}; (e) a finite undulatory swimmer swims faster (instantaneous body velocity and mean-squared polymer distention field are shown \cite{tfs10,tfsFig}); (f) and a thin helical body of arbitrary length can swim faster or slower, depending on the geometry and rotation rate~\cite{lpb11}. (g) The axial component of fluid velocity generated by a rotating, force-free helical filament is shown; a helical fluid volume external to the coil is carried upward with the translating body, while a helical fluid volume internal to the coil is shuttled downward.}
\label{Figure1}
\end{center}
\end{figure*}

The paper is organized as follows. We begin by presenting the mathematical model of a rotating, force-free helical body in an Oldroyd-B fluid, followed by a dimensional analysis and a description of the numerical method. Helices of small pitch angle are explored, for which we show a continuous departure of the small amplitude analytical theories from the results of the full model. We then turn to helices of large pitch angle where fluid elasticity is shown to increase the swimming speed to a local maximum at a Deborah number of order unity. Both helical waves and rigid body motions are considered, and the locomotive efficiency is addressed. We conclude with a summary and a discussion of future directions.

The experiments of Liu et al. suggest that the force-free swimming speed of a helical filament is broadly independent of its length~\cite{lpb11}. We therefore consider a right-handed helical body of infinite length with centerline $\b{x}(s,t)=b[\cos(k s+\Omega t)\b{\hat{x}}+\sin(ks+\Omega t)\b{\hat{y}}]+([1-(b k)^2]^{1/2} s+U^* t)\b{\hat{z}}$. Here, $b$ is the helical radius, $k$ is the wavenumber, $s$ is the arc-length, $\Omega$ is a fixed rotation rate, and $U^*$ is the swimming speed. The body is shaped such that the boundary in a cross-sectional plane perpendicular to the $\b{\hat{z}}$ axis is circular with radius $A/k$. The distinction between helical waves and rigid body rotation is a rotation of this circle about the centerline in the latter. 

At the length and velocity scales relevant for microorganisms, viscous effects dominate inertial effects~\cite{lp09} and the momentum and mass conservation equations are $\nabla p=\nabla \cdot \t{\tau}$ and $\nabla \cdot \b{u}=0$, where $p$ is the pressure and $\t{\tau}$ is the total deviatoric stress tensor. In the Oldroyd-B model, $\t{\tau}$ is the sum of a Newtonian solvent contribution, $\t{\tau}_s=\eta_s \t{\dot{\gamma}}$, and an extra polymeric contribution, $\t{\tau}_p$, where $\eta_s$ is the solvent viscosity, $\t{\dot{\gamma}}=\nabla\b{u}+\nabla\b{u}^T$ is the rate-of-strain tensor, and $\b{u}$ is the fluid velocity. Meanwhile, $\t{\tau}_p$ is described by an upper-convected Maxwell model in which a single elastic relaxation timescale, $\lambda_1$, and a viscous retardation timescale, $\lambda_2$, are introduced (with $\lambda_2=\lambda_1 \eta_s/\eta<\lambda_1$ and $\eta$ the total zero shear rate viscosity)~\cite{bah87,Larson99}. Scaling lengths on $1/k$, time on $1/\Omega$, velocities on $\Omega/k$, and stresses on $\eta \Omega$, the total deviatoric stress is found to satisfy the dimensionless constitutive relation:
\begin{gather}
\t{\tau}+De\, \ud{\t{\tau}}=\t{\dot{\gamma}}+(\eta_s/\eta)De\,\ud{\t{\dot{\gamma}}}.\label{constitutivep}
\end{gather}
Here we have defined the dimensionless Deborah number, $De=\lambda_1 \Omega$, which compares the body rotation rate to the rate of elastic relaxation. The upper convective time derivative in \eqref{constitutivep} is defined as $\ud{\t{\tau}}=\t{\tau}_t+\b{u}\cdot \nabla \t{\tau}-\nabla \b{u}^T\cdot \t{\tau}-\t{\tau}\cdot \nabla\b{u}$. More complex models include features such as multiple relaxation timescales and finite polymer extensibility \cite{bah87}.

\tcb{The Deborah numbers relevant to microorganism motility are likely to span a very wide range. Relaxation times of cervical mucus have been measured from $\lambda_1=0.03$ s to $\lambda_1=100$ s; spermatozoa in this environment undulate with frequencies between $20$ and $50$ Hz, corresponding to Deborah numbers that are $O(1)$ or much larger (see \cite{Lauga07,smith2009bend}). An intriguing example of Deborah number variation is exhibited in the swimming of \textit{H. pylori}, which rotates a helical flagellum at up to $5$ Hz in viscoelastic mucus with a relaxation time of approximately $100$ s, but releases pH-increasing urease enzymes that decreases the local relaxation time to nearly $0.05$ s, thereby decreasing the Deborah number from $O(1000)$ to $O(1)$ \cite{celli09}.} 

\begin{figure*}
\begin{center}
\hspace{-.2in}\includegraphics[width=6.5in]{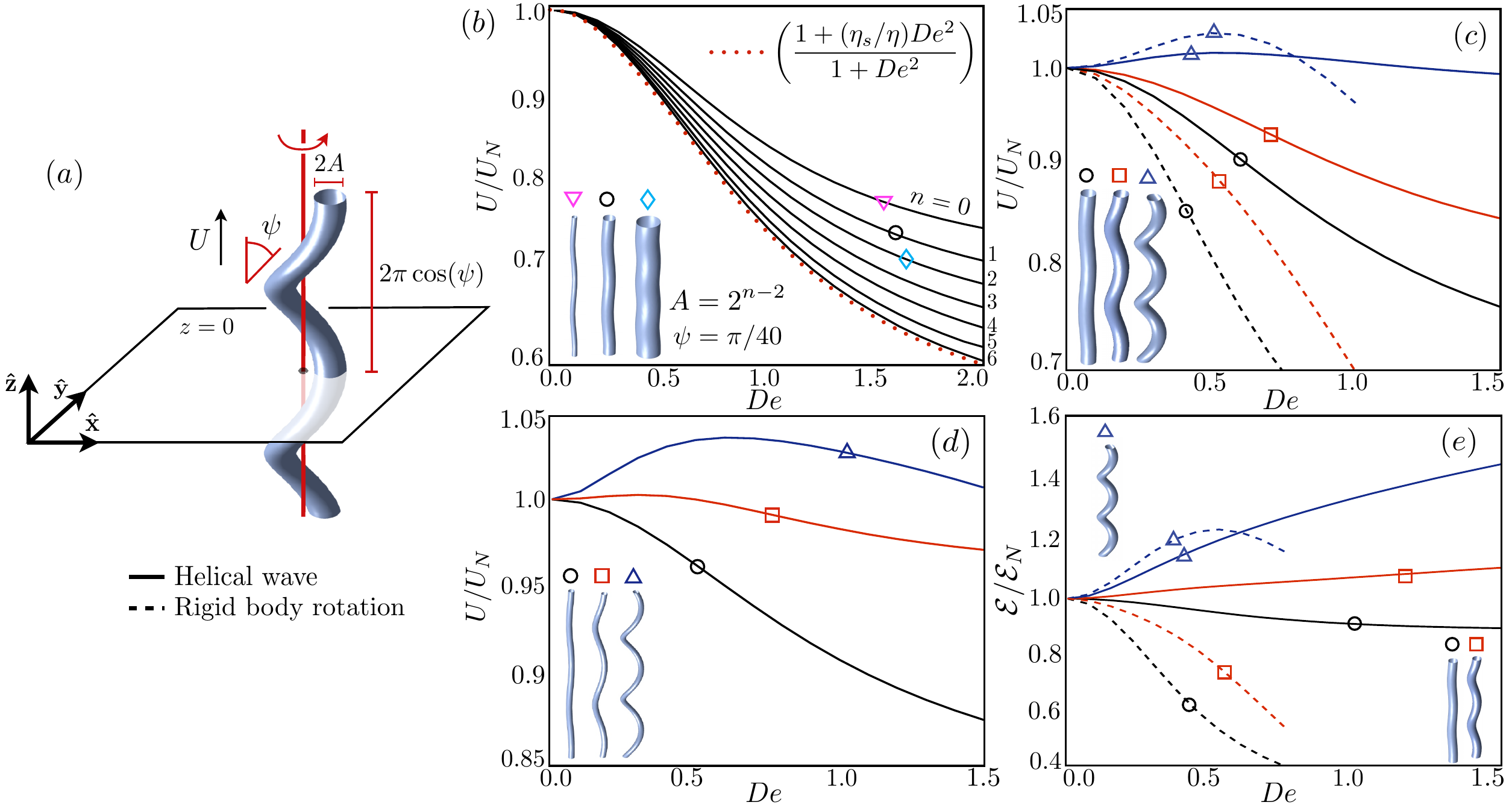}
\caption{(Color online) (a) Schematic of the dimensionless setup. The body rotates counter-clockwise with unit angular speed when seen from above and swims in the axial direction with dimensionless speed $U$. (b) Helical-wave swimming speed (normalized by the Newtonian swimming speed) of filaments of varying thickness, $A=2^{n-2}$. Here $\psi=\pi/40$ and $\eta_s/\eta=0.5$. Deviations from the theoretical result of Fu et al.~\cite{fwp09} (dotted line) are logarithmic in $A$. (c) Viscoelasticity leads to faster swimming for helices of sufficiently large pitch angle. The filament thickness is fixed, $A=0.5$. Solid lines denote helical waves, dashed lines denote rigid body rotation, and symbols denote pitch angle, $\psi=\pi/40,\,\pi/10$, and $\pi/5$. (d) Larger swimming speeds are achieved by thinner filaments of the same pitch angles as in (c); here $A=0.2$. (e) The normalized swimming efficiency for helical waves and rigid body rotations. Symbols denote the same helices as in (c).}
\label{Figure2}
\end{center}
\end{figure*}

Due to the interaction with the fluid, the helical filament translates along the axial direction with dimensionless speed $U$, as illustrated in Fig.~\ref{Figure2}a. A no-slip condition is assumed on the body surface, and for computational purposes we place the filament inside a very large container where we set $\b{u}=0$. The container is made sufficiently large so that further increases in its size have a negligible effect on the reported results. The problem is closed by requiring that the axial component of force on the body is zero. The constant swimming speed $U$ is determined by assuming that a locomotive steady state has been achieved and by exploiting helical symmetry: the flow and stress fields everywhere are given by translation and rotation of the flow field through the $z=0$ plane. Conversion to a helical coordinate system allows for $z$ derivatives to be written as planar derivatives on $z=0$. In a periodic steady state, time derivatives may be written as $z$ derivatives, and hence by planar derivatives. The Oldroyd-B equations are solved numerically using a mixed pseudo-spectral / finite differences approach. \tcb{The mathematical model and numerical method are described in greater detail in the supplementary material.}

We first compare the numerical results to the analytical work of Fu et al.~\cite{fwp09} in the case of a helical wave with small pitch angle. For $\psi=\pi/40$, the normalized swimming speed is shown in Fig.~\ref{Figure2}b for a range of Deborah numbers and filament sizes. Here as in the remainder of the paper we fix the viscosity ratio to $\eta_s/\eta=0.5$. Each solid line corresponds to a different filament radius, $A=2^{n-2}$, for $n=0,1,...,6$. By increasing the filament thickness the swimming speed converges monotonically to the analytical result, shown as a dashed line. Viscoelasticity in this case decreases the swimming speed of helices with small pitch angles, even for slender bodies, \tcb{contrasting with the enhanced speeds predicted in a viscous fluid in the presence of stationary obstacles \cite{Leshansky09}}. The departure of the results from the analytical theory are logarithmic in $A$, consistent with the analytical development \cite{fwp09}.

The analytical results at small pitch angle have been recovered, but can the increased swimming speeds seen in experiments be found? Figure~\ref{Figure2}c shows the normalized swimming speed as a function of the Deborah number for three different pitch angles, $\psi=\pi/40,\,\pi/10$, and $\pi/5$, with $A=0.5$. Helical wave and rigid body rotation results are shown as solid and dashed lines, respectively. For small Deborah numbers we observe $U/U_N=1+O(De^2)$, as required by symmetry. In both cases, for a given Deborah number in the regime considered, the swimming speed increases as the pitch angle is increased to $\psi=\pi/5$. Moreover, beyond a critical pitch angle we find a range of Deborah numbers for which the swimming speed is larger in a viscoelastic fluid than in a Newtonian fluid, just as observed in experiments~\cite{lpb11}. 

Rigid body motion, which generates an extra rotational flow around the helical filament as compared to helical waves, amplifies the effects of viscoelasticity, particularly for small pitch angles where rotational flow is dominant. Viscoelastic effects are amplified with decreasing $\eta_s/\eta$ as well (not shown). Filament thickness is also important; Fig.~\ref{Figure2}d shows the helical-wave swimming speeds of slender filaments ($A=0.2$), which are greater than those shown in Fig.~\ref{Figure2}c. In particular, for the intermediate pitch angle $\psi=\pi/10$, reducing the filament thickness introduces a regime in small Deborah number for which the relative swimming speed is greater than unity.

A microorganism may benefit by swimming with greater efficiency rather than greater speed. We evaluate a common measure of swimming efficiency, $\mathcal{E}=U^2/P$, where $P=(1/2)\int_{z=0} \t{\dot{\gamma}:\tau}\,dS$ is the rate of energy dissipation in the fluid per unit length~\cite{Lauga07,fwp09,tfs10}. The results are shown in Fig.~\ref{Figure2}e for the same helical shapes considered in Fig.~\ref{Figure2}c. An important distinction between the passage of helical waves and rigid body rotation is observed. For rigid body rotation, the work done on the fluid does not vary dramatically with Deborah number; just as was found for planar undulations, the swimming speed is a proxy for swimming efficiency in the case of rigid body rotation~\cite{tfs10}. The rotation of helices of large pitch angle therefore presents a means of swimming that is both faster and more efficient in a viscoelastic fluid when the rotation rate is properly tuned to the fluid environment. For helical waves, however, the swimming speed is not so clearly linked to the efficiency; for $\psi=\pi/10$ the relative swimming speed decreases with $De$ while the relative efficiency increases. By inspection of the no-slip boundary condition, the differences between rigid body rotation and helical waves are expected to diminish rapidly with both decreasing filament size and increasing pitch angle.

\begin{figure}[thbp]
\begin{center}
\hspace{0in}\includegraphics[width=3.4in]{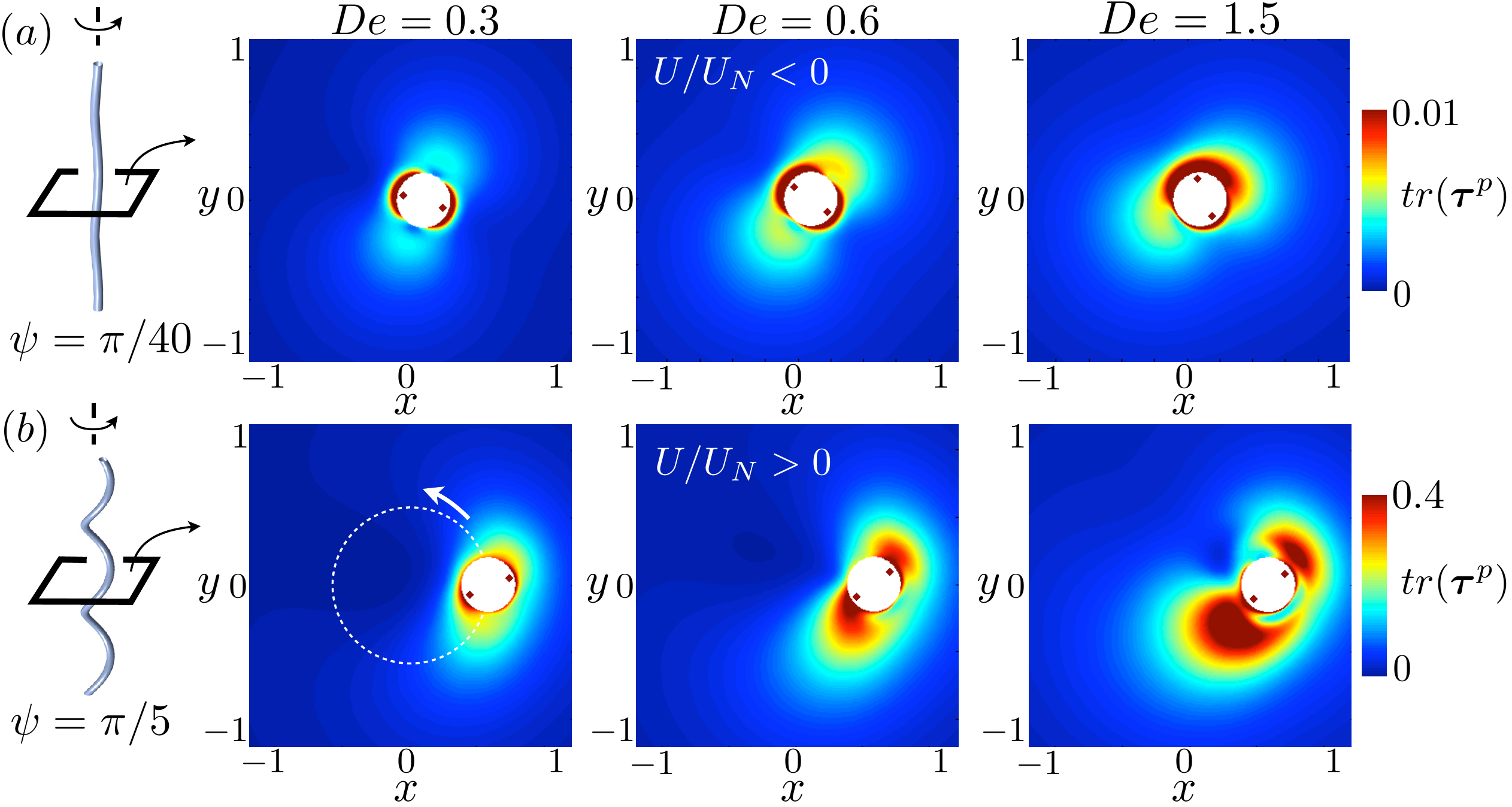}
\caption{(Color online) The mean-squared polymer distention fields, $tr(\t{\tau}_p)$, from the passage of helical waves are shown for three Deborah numbers, for filament thickness $A=0.2$ and pitch angles (a) $\psi=\pi/40$ and (b) $\psi=\pi/5$. The swimming speed decreases with Deborah number in (a) and increases in (b). The path of the body during rotation is indicated by a white dashed line in the bottom left panel. The largest polymer distention appears along the second/fourth quadrants in the former case, and along the first/third quadrants in the latter case, with the regions of maximal distention indicated by symbols.}
\label{Figure3}
\end{center}
\end{figure}

The effects of viscoelasticity on the swimming of helices are not easily predicted by thought experiments. The flow field created by a rotating helix in a Newtonian fluid is intricate; the extra polymeric stresses that develop due to this flow field, the response of the flow field to the polymeric stresses, and the interaction of solvent and polymeric forces with the helix all provide for a complex and highly nonlinear system. 

We do, however, observe a distinction in the polymeric stresses between cases where viscoelasticity either reduces or increases the swimming speed. Consider the spatial distribution of $tr(\t{\tau}_p)$ for the passage of helical waves along two different helical geometries. This quantity measures the mean-squared distention of the elastic polymers, and is shown for helices of small and large pitch angle in Fig.~\ref{Figure3}. For small Deborah numbers, the polymers are stretched most in regions of large fluid shear, which are at the leftmost and rightmost points of the circular boundaries in Fig.~\ref{Figure3} (the inner and outer edges of the filament, respectively). Fluid shear is largest in these regions due not only to the motion of the filament through the $z=0$ plane, but also to the arrangement of the axial fluid velocity, as shown for $De=0$ in \tcb{Fig.~\ref{Figure1}g}.

Increasing the Deborah number, however, shows a distinct difference in the organization of polymeric stress in these two cases. For the helix of smaller pitch angle the primary regions of extra stress rotate clockwise. As these regions are displaced, they affect the underlying flow field, and both conspire to reduce the filament swimming speed. For the helix of larger pitch angle, however, we observe the opposite shift: extra polymeric stresses have shifted counter-clockwise. Much as in the case of flow past a cylinder, the reorganization of elastic fluid stress acts to shift the distribution of pressure, which further contributes to adjustments in both vertical and horizontal fluid forces on the body~\cite{pd99}. 

Recall that a Deborah number of order unity indicates a helical rotation rate comparable to the rate of elastic relaxation. When $De\approx1$, polymers distended by the motion of the body release stored elastic energy on a special timescale. Namely, a timescale such that the body can revisit the viscoelastic wake created on each pass through the same location. For helices of small pitch angle, the motion of the body in the $z=0$ plane is muted, and fluid parcels do not make large excursions in the plane. With a larger pitch angle, the filament travels on a much wider circuit relative to its thickness, the body can interact with its own viscoelastic wake, and elastic stresses can be transmitted from the wake back onto the body upon its return. This heuristic also suggests that a smaller filament thickness (with pitch angle fixed) may lead to increased swimming speeds, as we have already observed in Fig.~\ref{Figure2}d.

We have shown that the introduction of viscoelasticity can either decrease or increase the force-free swimming speed and swimming efficiency of a rotating helical filament, depending on the helical geometry, the material properties of the fluid, and the body rotation rate. The results of our investigation connect the small amplitude theories, experimental observations, and numerical investigations of the biological and mechanical experiments shown in Fig.~\ref{Figure1}. \tcb{Our findings may add context to the recent discovery that \textit{H. pylori} reduces mucus elasticity to a range more suitable for effective locomotion \cite{celli09}.} Future work will explore the effects of viscoelasticity on the propulsion of elastic helical filaments, on flagellar bundling, and on more intricate solid structures such as those observed in bacterial polymorphism~\cite{Calladine75,sp05,sl11}. Similar behaviors are expected in a nearby fluid pumping problem, where fluid is transported downward either faster or slower than the same in a Newtonian setting.

We acknowledge helpful discussions with M. Graham, M. Shelley, R. Ewoldt, and G. Forest, and we are especially grateful to K. Breuer for ongoing collaboration. This work was supported by the NSF Grant No. CBET-0854108.

\bibliographystyle{unsrt}
\bibliography{Bigbib}
\end{document}